\documentclass[12pt,a4paper]{article}

\usepackage[T1]{fontenc}
\usepackage[utf8]{inputenc}
\usepackage{amsmath}
\usepackage{amssymb}
\usepackage{hyperref}
\usepackage[final]{graphicx}
\usepackage{bbm,latexsym,epsfig}

\newcommand{\be}{\begin{equation}}
\newcommand{\ee}{\end{equation}}
\newcommand{\ba}{\begin{eqnarray}}
\newcommand{\ea}{\end{eqnarray}}
\newcommand{\bs}{\begin{subequations}}
\newcommand{\es}{\end{subequations}}
\newcommand{\no}{\nonumber \\}

\begin{document}

\title{
\normalsize \hfill  \\
\normalsize \hfill CFTP/13-028
\\[8mm]
\LARGE Golden ratio lepton mixing and \\ nonzero reactor angle with $A_5$}

\author{
I.~de Medeiros Varzielas$^{(1)}$\thanks{E-mail: {\tt ivo.de@unibas.ch}}
\ and
L.~Lavoura$^{(2)}$\thanks{E-mail: {\tt balio@cftp.ist.utl.pt}}
\\*[8mm]
$^{(1)} \!$
\small Department of Physics, University of Basel,
\\
\small Klingelbergstra\ss e 82, CH-4056 Basel, Switzerland
\\*[4mm]
$^{(2)} \!$
\small Universidade de Lisboa, Instituto Superior T\'ecnico, CFTP,
\\
\small 1049-001 Lisboa, Portugal
\\*[7mm]
}

\date{\today}

\maketitle

\begin{abstract}
We furnish a supersymmetric extension of the Standard Model
with a flavour discrete symmetry $A_5$
under which the lepton fields transform as an irreducible triplet.
Additional (`flavon') superfields are used to break $A_5$ into
a $\mathbb{Z}_2 \times \mathbb{Z}_2$ subgroup in the charged-lepton sector
and another $\mathbb{Z}_2$ subgroup in the neutrino sector.
The first column of the resulting lepton mixing matrix is predicted
and has entries which are related to the golden ratio.
Using the observed $\theta_{13}$ as input,
our model predicts a solar mixing angle $\theta_{12}$
in very good agreement with experiment;
it also predicts a correlation
between the atmospheric mixing angle $\theta_{23}$
and the $CP$-violating Dirac phase $\delta$.
\end{abstract}

\section{Introduction}

The experimental measurement
of the mixing angle $\theta_{13}$~\cite{minos,chooz,dayabay,reno}
has been of great importance in advancing our understanding of lepton mixing.
It has ruled out
(at least at leading order)
many mixing schemes which have $\theta_{13}=0$;
two of them are the golden ratio\footnote{The golden ratio is the number
$\varphi \equiv \left. \left( 1 + \sqrt{5} \right) \right/ 2 \approx 1.618$.
For its importance in mathematics and the arts,
see for instance {\tt http://en.wikipedia.org/wiki/Golden-ratio}.}
schemes GR1~\cite{Datta:2003qg, Kajiyama:2007gx, Everett:2008et,
Chen:2010ty, Feruglio:2011qq}
and GR2~\cite{Rodejohann:2008ir, Adulpravitchai:2009bg}
(see also refs.~\cite{Albright:2010ap, Ding:2011cm, Cooper:2012bd}).
Those two schemes make predictions for the solar mixing angle $\theta_{12}$;
in GR1 $\cot{\theta_{12}} = \varphi$
and in GR2 $\cos{\theta_{12}} = \varphi / 2$.
However,
both those schemes also predict $\theta_{13} = 0$---hence
a spurious $CP$-violating phase $\delta$---and
maximal atmospheric mixing angle $\theta_{23}$;
the first of these additional predictions
is now excluded by the experimental observation
of a nonzero $\theta_{13}$~\cite{minos,chooz,dayabay,reno}.

GR1 and GR2 were often implemented in models
by using either the discrete group $A_5$
or its double cover~\cite{Everett:2010rd, Hashimoto:2011tn, Kajiyama:2013wwa}.
In this paper we propose a model based on $A_5$
which leads to \emph{an alternative golden ratio mixing scheme}\/
which is in full agreement with the experimental data.
In our scheme,
just as in GR2,
$\cos{\theta_{12}} = \varphi / 2$ if $\theta_{13}$ vanishes;
but,
in our scheme $\theta_{13}$ is not necessarily zero,
since our model only predicts the first column of the lepton mixing matrix.
Our model has the additional feature that
it is not just \emph{one}\/ matrix element
of the lepton mixing matrix $U$
which is related to the golden ratio---our model
predicts both $|U_{e1}| = \varphi / 2$
\emph{and}\/ $|U_{\mu 1}| = 1 \left/ \left( 2 \varphi \right)
\right.$.\footnote{An
alternative version of our model has
$U_{\tau 1} = 1 \left/ \left( 2 \varphi \right) \right.$ instead.
The two versions are related by the approximate $\mu$--$\tau$
interchange symmetry which seems to hold in $U$.}
In contrast to GR1 and GR2,
and akin to trimaximal mixing~\cite{grimus}
and other schemes which only fix one column of $U$
(see for instance refs.~\cite{Rodejohann:2012cf, Varzielas:2012pa}
and the references therein),
our model does not fix (at leading order) all three mixing angles,
but instead just predicts two relations among them.

Like many other previous models for lepton mixing,
ours is based on the paradigm of a flavour symmetry group
$\mathcal{G}$---in our case,
$A_5$---which is broken to
an Abelian subgroup $\mathcal{F}_\ell$ in the charged-lepton sector
and to another Abelian subgroup $\mathcal{F}_\nu$ in the neutrino sector.
Various possibilities for $\mathcal{G}$
and for its subgroups $\mathcal{F}_\ell$ and $\mathcal{F}_\nu$
have been searched systematically before~\cite{lam,hernandez,lindner} (see also \cite{Hanlon:2013ska}).
However,
to the best of our knowledge all those searches restricted $\mathcal{F}_\ell$
to be a $\mathbbm{Z}_n$ group with $n \ge 3$.
Additionally,
some works~\cite{deAdelhartToorop:2011re, Feruglio:2012cw, Hagedorn:2013nra}
have considered the possibility that
both $\mathcal{F}_\ell$ and $\mathcal{F}_\nu$ are
(distinct)
$\mathbbm{Z}_2 \times \mathbbm{Z}_2$ subgroups of $\mathcal{G}$.
Of particular relevance for us is ref.~\cite{deAdelhartToorop:2011re},
where $\mathcal{G} = A_5$ has been considered;
in that work a mixing scheme has been suggested
in which $\theta_{13}$ has a specific non-zero value
related to the golden ratio but larger than the phenomenological value.
In the present paper,
$\mathcal{F}_\ell$ is $\mathbbm{Z}_2 \times \mathbbm{Z}_2$
and $\mathcal{F}_\nu$ is $\mathbbm{Z}_2$;
this possibility has never been considered before
and for this reason our model apparently evaded
all the searches previously made.

The outline of this paper is as follows.
In section~\ref{sec:model} we present our model.
In section~\ref{sec:fit} we fit the predictions of our model
to the phenomenological data.
A summary is presented in section~\ref{sec:conclusions}.
An appendix is devoted to some facts on the group $A_5$.

\section{The model}
\label{sec:model}

The group $A_5$ is the \textit{alternating group}\/
of the even permutations of five objects.
It has $5! / 2 = 60$ elements
and it is isomorphic to the symmetry group
of the platonic solids icosahedron and dodecahedron;
we also call it the \textit{icosahedral group}\/ (IG).

There are various possible presentations of the IG.
In this paper we use a presentation~\cite{dickson,grimusludl}
wherein $A_5$ is the group generated by three transformations $a$,
$b$,
and $c$ which satisfy
\be
\label{presentation}
a^2 = b^3 = c^2
= \left( b a \right)^3 = \left( a c \right)^3 = \left( b c \right)^2 = e,
\ee
where $e$ is the identity transformation.

The IG has five $\mathbb{Z}_2 \times \mathbb{Z}_2$
subgroups.\footnote{All the subgroups of $A_5$
are listed in ref.~\cite{Ding:2011cm},
but in a presentation different from ours.}
One of them is constituted by $e$,
$a$,
$b a b^2$,
and $b^2 a b$.
The IG also has fifteen $\mathbb{Z}_2$ subgroups;
one of them is formed by $e$ and $c$.
Those are the subgroups of the IG that are crucial in this paper.

The IG has five inequivalent irreducible representations (irreps):
the $\mathbf{1}$ (trivial representation),
the $\mathbf{3}_1$,
the $\mathbf{3}_2$,
the $\mathbf{4}$,
and the $\mathbf{5}$,
where the boldface numbers are the dimensions of the irreps.
These are all real representations.
In this paper we shall only use the $\mathbf{1}$,
the $\mathbf{3}_1$,
and the $\mathbf{5}$;
they are given,
in convenient bases,
in appendix~\ref{apa}.

Let $D_{Lj}$ ($j = 1, 2, 3$) be the superfields containing
the gauge-$SU(2)$ doublets of left-handed leptons.
We place the three $D_{Lj}$ in a $\mathbf{3}_1$ of $A_5$;
let $D_L$ denote it.
Similarly,
the right-handed charged leptons,
which are singlets of the gauge $SU(2)$,
are placed in a $\mathbf{3}_1$ of $A_5$;
let $\ell_R$ denote it.
Our supersymmetric model\footnote{In our model supersymmetry is needed
only in order to implement in a consistent fashion
the required alignments of vacuum expectation values.
In particular,
supersymmetry is useful to separate the flavons and driving fields
operating in the charged-lepton sector
from the flavons and driving fields appearing in the neutrino sector,
in such a way that the alignments may differ from each other
in those two sectors.}
has an $R$-symmetry under which
both $D_L$ and $\ell_R$ have $R$-charge $+1$;
allowed superpotential terms must have $R$-charge $+2$. 

The neutrino masses originate from a type-II seesaw mechanism,
triggered by a gauge-$SU(2)$ triplet $T$
which is invariant under $A_5$.
In order to achieve $A_5$-invariance,
we introduce \textit{flavon}\/ superfields $S$ and $P$
which are gauge-invariant and transform as a $\mathbf{1}$ and a $\mathbf{5}$,
respectively,
of $A_5$.
Both the flavons and $T$ have zero $R$-charge.
The term of the (non-renormalizable) superpotential
relevant for the neutrino masses then is
\be
\label{numass}
\frac{1}{\Lambda}\, D_L T D_L \left( z_S S + z_P P \right),
\ee
where $z_{S,P}$ are numerical coefficients
and $\Lambda$ is the high energy (cutoff) scale
which suppresses all the non-renormalizable
terms.\footnote{Note that the origin of $\Lambda$ can only be specified
through a higher-scale completion of the model,
wherein $\Lambda$ might naturally be identified
as the mass of some messenger superfield charged under the flavour symmetry
(see \textit{e.g.}\ refs.~\cite{Varzielas:2010mp,Varzielas:2012ai}).
At the level that we consider here
it is not possible to relate $\Lambda$ to the scale of $SU(2)$ breaking.
Indeed,
the neutrino mass scale depends on the ratio $\langle S \rangle / \Lambda$
rather that on the absolute scales $\langle S \rangle$ and $\Lambda$---which
can easily be several orders of magnitude higher than the Fermi scale.}
We note that the product of the two $D_L$ in eq.~(\ref{numass})
is \emph{symmetric},
therefore,
because of eq.~(\ref{31product}),
it can only contain the representations $\mathbf{1}$ and $\mathbf{5}$ of $A_5$.

Analogously,
the charged-lepton masses originate in the superpotential term
\be
\frac{1}{\Lambda}\, D_L H \ell_R
\left( \bar z_S \bar S + \bar z_P \bar P \right),
\ee
where $H$ is an $A_5$-invariant $SU(2)$ doublet
while $\bar S$ and $\bar P$ are a $\mathbf{1}$ and a $\mathbf{5}$ of $A_5$,
respectively,
which are invariant under the gauge group.\footnote{According
to eq.~(\ref{31product}),
an additional flavon in a $\mathbf{3}_1$ of $A_5$
might also contribute to the charged-lepton masses.
We dispense with it because it is superfluous.}
The $\bar z_{S,P}$ are numerical coefficients.

\subsection{Neutrino mass matrix \label{sec:neutrino}}

Let $s$ denote the vacuum expectation value (VEV) of $S$
and $\left( p_1,\, p_2,\, p_3,\, p_4,\, p_5 \right)$ denote the VEV of $P$.
Then the neutrino Majorana mass matrix following from eq.~(\ref{numass}) is,
according to eq.~(\ref{see31product}),
given by
\bs
\label{mnu}
\ba
\left( M_\nu \right)_{11} &=&
z_S s + z_P \left( - \frac{\mu_+}{2}\, p_4
+ \frac{1 - \mu_-}{2 \sqrt{3}}\, p_5 \right),
\\
\left( M_\nu \right)_{22} &=&
z_S s + z_P \left( - \frac{\mu_-}{2}\, p_4
+ \frac{\mu_+ - 1}{2 \sqrt{3}}\, p_5 \right),
\\
\left( M_\nu \right)_{33} &=&
z_S s + z_P \left( - \frac{1}{2}\, p_4
+ \frac{\mu_- - \mu_+}{2 \sqrt{3}}\, p_5 \right),
\\
\left( M_\nu \right)_{12} = \left( M_\nu \right)_{21} &=&
\frac{z_P p_1}{\sqrt{2}}\, ,
\\
\left( M_\nu \right)_{13} = \left( M_\nu \right)_{31} &=&
\frac{z_P p_3}{\sqrt{2}}\, ,
\\
\left( M_\nu \right)_{23} = \left( M_\nu \right)_{32} &=&
\frac{z_P p_2}{\sqrt{2}}\, ,
\ea
\es
where
\be
\mu_\pm \equiv \frac{\pm \sqrt{5} -1}{2}\, .
\ee
These numbers satisfy
\bs
\ba
1 + \mu_+ + \mu_- &=& 0,
\\
\mu_+ \mu_ - &=& -1,
\\
\mu_+^2 + \mu_-^2 &=& 3.
\ea
\es
Obviously,
$\mu_- = - \varphi$ and $\mu_+ = 1 / \varphi$.

We assume that
\emph{the VEV of $P$ preserves the $\mathbb{Z}_2$ subgroup of $A_5$
generated by $c$.}
According to eq.~(\ref{c5}),
this means that
\be
\label{PVEV}
\left( \begin{array}{c} p_1 \\ p_2 \\ p_3 \\ p_4 \\ p_5 \end{array} \right) =
\left( \begin{array}{c}
x + 2 z \\
x + \sqrt{6}\, y - z \\
x - \sqrt{6}\, y - z\\
3 \sqrt{2} z \\
2 y \\
\end{array} \right).
\ee
It follows from eqs.~(\ref{mnu})--(\ref{PVEV}) that
\be
\label{eigenvector}
M_\nu v
=
\left[ z_S s + z_P \left( - \frac{x}{\sqrt{2}}
+ 2 \sqrt{\frac{5}{3}}\, y - \sqrt{2} z \right) \right] v,
\ee
where
\be
\label{vmu}
v \equiv \frac{1}{2}
\left( \begin{array}{c} \mu_- \\ \mu_+ \\ 1 \end{array} \right).
\ee
This means that $v$ is a (normalized) eigenvector of $M_\nu$.
This is a consequence of the fact that,
because the VEV of $P$ preserves $c$,
\be
C_3 M_\nu C_3 = M_\nu,
\ee
where $C_3$ is the matrix that represents $c$
in the representation $\mathbf{3}_1$ of $A_5$.
Now,
$C_3$ has a non-degenerate eigenvalue $+1$
and $v$ is the eigenvector corresponding to that eigenvalue;
therefore,
$v$ must also be an eigenvector of $M_\nu$:
\be
\left( C_3 M_\nu C_3 = M_\nu,\ C_3 v = v \right)
\Rightarrow
M_\nu v = C_3 M_\nu v \Rightarrow M_\nu v \propto v.
\ee

The neutrino masses are free in our model,
\textit{i.e.}\ the model does not purport to either predict or constrain them.
Their overall scale is fixed by $z_S s$
and the differences among them depend on $z_P x$,
$z_P y$,
and $z_P z$.
These four parameters are all free.

\subsection{Charged-lepton mass matrix}

Let $\bar s$ denote the VEV of $\bar S$
and $\left( \bar p_1,\, \bar p_2,\, \bar p_3,\, \bar p_4,\, \bar p_5 \right)$
denote the VEV of $\bar P$.
Then,
the charged-lepton mass matrix $M_\ell$
has exactly the same form as the neutrino mass matrix in eqs.~(\ref{mnu}):
\bs
\label{mell0}
\ba
\left( M_\ell \right)_{11} &=&
\bar z_S \bar s + \bar z_P \left( - \frac{\mu_+}{2}\, \bar p_4
+ \frac{1 - \mu_-}{2 \sqrt{3}}\, \bar p_5 \right),
\\
\left( M_\ell \right)_{22} &=&
\bar z_S \bar s + \bar z_P \left( - \frac{\mu_-}{2}\, \bar p_4
+ \frac{\mu_+ - 1}{2 \sqrt{3}}\, \bar p_5 \right),
\\
\left( M_\ell \right)_{33} &=&
\bar z_S \bar s + \bar z_P \left( - \frac{1}{2}\, \bar p_4
+ \frac{\mu_- - \mu_+}{2 \sqrt{3}}\, \bar p_5 \right),
\\
\left( M_\ell \right)_{12} = \left( M_\ell \right)_{21} &=&
\frac{\bar z_P \bar p_1}{\sqrt{2}}\, ,
\\
\left( M_\ell \right)_{13} = \left( M_\ell \right)_{31} &=&
\frac{\bar z_P \bar p_3}{\sqrt{2}}\, ,
\\
\left( M_\ell \right)_{23} = \left( M_\ell \right)_{32} &=&
\frac{\bar z_P \bar p_2}{\sqrt{2}}\, .
\ea
\es
We assume that \emph{the VEV of $\bar P$ is invariant under
the $\mathbb{Z}_2 \times \mathbb{Z}_2$ subgroup of $A_5$
generated by $b^2 a b$ and $b a b^2$}.
Since in the $\mathbf{5}$ of $A_5$
\bs
\ba
b a b^2 &\to& \mathrm{diag} \left( -1,\, -1,\, 1,\, 1,\, 1 \right),
\\
b^2 a b &\to& \mathrm{diag} \left( -1,\, 1,\, -1,\, 1,\, 1 \right),
\ea
\es
this means that $\bar p_1 = \bar p_2 = \bar p_3 = 0$.
The charged-lepton mass matrix is then diagonal and
one possible assignment of the charged-lepton masses is
\bs
\label{mell}
\ba
m_e &=& \left| \bar z_s \bar s - \frac{\bar z_P}{2} \left( \bar p_4 \mu_+ +
\frac{\mu_- - 1}{\sqrt{3}}\, \bar p_5 \right) \right|,
\\
m_\mu &=& \left| \bar z_s \bar s - \frac{\bar z_P}{2} \left( \bar p_4 \mu_- +
\frac{1 - \mu_+}{\sqrt{3}}\, \bar p_5 \right) \right|,
\\
m_\tau &=& \left| \bar z_s \bar s - \frac{\bar z_P}{2} \left( \bar p_4 +
\frac{\mu_+ - \mu_-}{\sqrt{3}}\, \bar p_5 \right) \right|.
\ea
\es

\subsection{Driving fields}

In order to align the VEV of $P$ along the directions in eq.~(\ref{PVEV}),
we introduce a `driving' superfield $\Delta$,
which is a $\mathbf{5}$ of $A_5$ and has $R$-charge $+2$.
The relevant (renormalizable) interaction terms are then
\be
\label{Deltaalign}
\Delta \left[
a_1 S P + a_2 \left( P P \right)_1 + a_3 \left( P P \right)_2
\right], 
\ee
where we have taken into account that there are two independent $\mathbf{5}$
in the product of two $\mathbf{5}$ of $A_5$,
see eq.~(\ref{5product}).
We use for those two $\mathbf{5}$ the convention explicit
in eqs.~(\ref{see5product}) and thereby identify
two separate coefficients $a_2$ and $a_3$ in eq.~(\ref{Deltaalign}).
The minimization of the terms in eq.~(\ref{Deltaalign}) implies
\bs
\label{VEValign}
\ba
\label{al1}
a_1 s p_1
+ \sqrt{2}\, a_2 p_1 p_5
+ \sqrt{\frac{2}{3}}\, a_3 \left( p_1 p_4 + \sqrt{2}\, p_2 p_3 \right) &=& 0,
\\
\label{al2}
a_1 s p_2
+ \frac{a_2}{\sqrt{2}} \left( - p_2 p_5 + \sqrt{3}\, p_2 p_4 \right) & &
\no
+ \frac{a_3}{\sqrt{6}} \left( - p_2 p_4 - \sqrt{3}\, p_2 p_5
+ 2 \sqrt{2}\, p_1 p_3 \right) &=& 0,
\\
\label{al3}
a_1 s p_3
+ \frac{a_2}{\sqrt{2}} \left( - p_3 p_5 - \sqrt{3}\, p_3 p_4 \right) & &
\no
+ \frac{a_3}{\sqrt{6}} \left( - p_3 p_4 + \sqrt{3}\, p_3 p_5
+ 2 \sqrt{2}\, p_1 p_2 \right) &=& 0,
\\
\label{al4}
a_1 s p_4
+ \frac{a_2}{2 \sqrt{2}} \left( - 2 p_4 p_5 + \sqrt{3}\, p_2^2
- \sqrt{3}\, p_3^2 \right) & &
\no
+ \frac{a_3}{2 \sqrt{6}} \left( 2 p_1^2 - p_2^2 - p_3^2
+ 3 p_4^2 - 3 p_5^2 \right) &=& 0,
\\
\label{al5}
a_1 s p_5
+ \frac{a_2}{2 \sqrt{2}} \left( 2 p_1^2 - p_2^2 - p_3^2
- p_4^2 + p_5^2  \right) & &
\no
+ \frac{a_3}{2 \sqrt{6}} \left( \sqrt{3}\, p_3^2 - \sqrt{3}\, p_2^2
- 6 p_4 p_5 \right) &=& 0.
\ea
\es
Equations~(\ref{VEValign}) are (partially) solved in a self-consistent way,
which means that one assumes a form for the VEV of $P$
and checks that that form is compatible with eqs.~(\ref{VEValign}).
In practice,
this just implies that the assumed form of the VEV of $P$
should preserve some subgroup of $A_5$;
in our case that subgroup is the one formed by $e$ and $c$.
Thus,
inserting eq.~(\ref{PVEV}) into eqs.~(\ref{VEValign}),
one obtains
\bs
\label{xyz}
\ba
a_1 s x + 8 \sqrt{2} a_2 y z
+ \frac{2 a_3}{\sqrt{3}} \left( x^2 + 2 z^2 - 4 y^2 \right) &=& 0,
\\
a_1 s y + \frac{a_2}{\sqrt{2}} \left( 3 x z - 3 z^2 - 2 y^2 \right)
- \sqrt{3} a_3 y \left( x + 2 z \right) &=& 0,
\\
a_1 s z + \sqrt{2} a_2 y \left( x - 2 z \right)
+ \frac{a_3}{\sqrt{3}} \left( x z + 5 z^2 - 2 y^2 \right) &=& 0.
\ea
\es
Thus,
eq.~(\ref{PVEV}) constitutes a solution to eqs.~(\ref{VEValign})
provided $x$,
$y$,
and $z$ satisfy eqs.~(\ref{xyz}).
In practice,
given $a_{1,2,3}$ and $s$,
one may in principle solve eqs.~(\ref{VEValign}) to determine (nonzero) $x$,
$y$,
and $z$.

In an analogous fashion,
in order to align the VEV of $\bar P$
we use a driving superfield $\bar \Delta$
which is a $\mathbf{5}$ of $A_5$.
The relevant interaction terms are
\be
\label{barDeltaalign}
\bar \Delta \left[
\bar a_1 \bar S \bar P + \bar a_2 \left( \bar P \bar P \right)_1
+ \bar a_3 \left( \bar P \bar P \right)_2
\right], 
\ee
just as in eq.~(\ref{Deltaalign}).
The corresponding minimization equations are analogous to eqs.~(\ref{VEValign}).
However,
now we make the self-consistent assumption
$\bar p_1 = \bar p_2 = \bar p_3 = 0$,
and this renders identically zero
the equations corresponding to eqs.~(\ref{al1})--(\ref{al3});
the only remaining equations are the ones analogous
to eqs.~(\ref{al4}) and~(\ref{al5}):
\bs
\label{al6}
\ba
\bar a_1 \bar s \bar p_4
- \frac{\bar a_2}{\sqrt{2}}\, \bar p_4 \bar p_5
+ \frac{3 \bar a_3}{2 \sqrt{6}} \left( \bar p_4^2 - \bar p_5^2 \right) &=& 0,
\\
\bar a_1 \bar s \bar p_5
- \frac{\bar a_2}{2 \sqrt{2}} \left( \bar p_4^2 - \bar p_5^2  \right)
+ \frac{\sqrt{6}\, \bar a_3}{2}\, \bar p_4 \bar p_5 &=& 0.
\ea
\es
These two equations in principle determine $\bar p_4$ and $\bar p_5$
as functions of $\bar s$ and of the coefficients $\bar a_1$,
$\bar a_2$,
and $\bar a_3$.
There is sufficient freedom to fit the lepton masses
to their experimental values.\footnote{In this work
we do not attempt to address the problem
of the hierarchy of the charged-lepton masses.}

\subsection{Additional symmetries}

An additional symmetry is needed
that separates the flavons and driving fields
responsible for the neutrino mass terms
from the ones responsible for the charged-lepton mass terms.
We choose for this effect a $\mathbbm{Z}_3 \times \mathbbm{Z}_3$ symmetry.
Table~\ref{nuZ3} lists the charges of the superfields
under all the symmetries of the model.
\begin{table}[ht]
\centering
\begin{tabular}{|c|cc|cc|cccc|cc|}
Superfield & $D_L$ & $\ell_R$ & $H$ & $T$ & $S$ & $P$ & $\bar S$ & $\bar P$ &
$\Delta$ & $\bar \Delta$
\\ \hline
$A_5$ & $\mathbf{3}_1$ & $\mathbf{3}_1$ & $\mathbf{1}$ & $\mathbf{1}$ &
$\mathbf{1}$ & $\mathbf{5}$ & $\mathbf{1}$ & $\mathbf{5}$ &
$\mathbf{5}$ & $\mathbf{5}$ \\
$\mathbbm{Z}_{3\nu}$ & 1 & 2 & 0 & 0 & 1 & 1 & 0 & 0 & 1 & 0 \\
$\mathbbm{Z}_{3\ell}$ & 0 & 2 & 0 & 0 & 0 & 0 & 1 & 1 & 0 & 1 \\
$U(1)_R$ & $1$ & $1$ & $0$ & $0$ & $0$ & $0$ & $0$ & $0$ & $2$ & $2$ \\
\end{tabular} 
\caption{The superfields and their transformation properties
under the symmetries of the model. \label{nuZ3}}
\end{table}

Note that,
as the superfields $H$ and $T$ are invariant under the additional symmetries,
the model may be embedded in a full supersymmetric type II seesaw construction.
This requires two gauge-$SU(2)$ doublets
and two triplets~\cite{Hambye:2000ui, Rossi:2002zb}.
Here we show only $H$ and $T$ for simplicity.

\section{Predictions for the mixing angles}
\label{sec:fit}

It is clear at this stage that
the solutions that we assumed for the VEVs are not unique.
Indeed,
we have assumed,
in the neutrino and charged-lepton sectors,
different self-consistent solutions
for entirely analogous sets of alignment equations.
Nonetheless,
the set of VEV directions that solve those alignment constraints
is discrete and,
thus,
it is reasonable to assume our desired alignments as the outcome.
This assumption is justified
by the phenomenological consequences
of our selected VEVs of $P$ and $\bar{P}$,
which we describe in this section.

\begin{figure}
\begin{center}
\epsfig{file=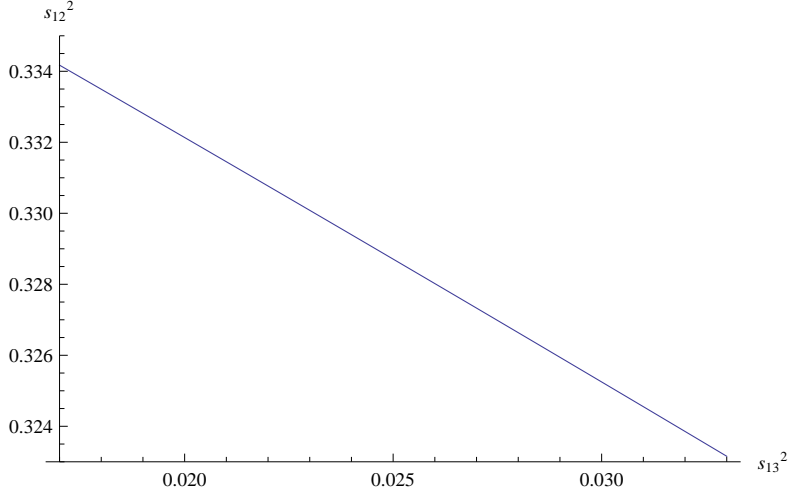,width=0.75\textwidth}
\end{center}
\caption{$s_{12}^2$ as a function of $s_{13}^2$
within the $3 \sigma$ phenomenological range
$s_{13}^2 \in \left[ 0.017,\, 0.033 \right].$
\label{fig:s12sq}}
\end{figure}

We have found that,
with our choice of VEVs,
the charged-lepton mass matrix is diagonal
while the neutrino mass matrix
has one eigenvector proportional to $\left( \mu_{-},\, \mu_{+},\, 1 \right)^T$.
This eigenvector is therefore one of the columns of the lepton mixing matrix
\be
\label{eq:stdpar}
U = \left( \begin{array}{ccc}
c_{12} c_{13} &
s_{12} c_{13} &
s_{13} e^{- i \delta} \\
- s_{12} c_{23} - c_{12} s_{23} s_{13} e^{i \delta} &
c_{12} c_{23} - s_{12} s_{23} s_{13} e^{i \delta} &
s_{23} c_{13} \\
s_{12} s_{23} - c_{12} c_{23} s_{13} e^{i \delta} &
- c_{12} s_{23} - s_{12} c_{23} s_{13} e^{i \delta} &
c_{23} c_{13}
\end{array} \right) P,
\ee
where $c_i \equiv \cos{\theta_i}$ and $s_i \equiv \sin{\theta_i}$
for $i = 12, 13, 23$ and $P$ is a diagonal unitary matrix
containing the so-called Majorana phases.
We now choose to identify this column as being the first one and,
moreover,
we take the charged leptons in eq.~(\ref{mell}) in the standard ordering.
We thus obtain
\bs
\ba
\label{pred1}
c_{12}^2 c_{13}^2 &=& \frac{\mu_-^2}{4},
\\
\label{pred2}
\left( c_{23}^2 - s_{23}^2 \right) \left( s_{12}^2 - c_{12}^2 s_{13}^2 \right)
+ 4 c_{12} s_{12} c_{23} s_{23} s_{13} \cos{\delta} &=& \frac{\mu_+^2 - 1}{4}.
\ea
\es
We obtain from eq.~(\ref{pred1}) that
\be
\label{s12sq}
s_{12}^2
= 1 - \frac{\mu_{-}^2}{4 \left( 1 - s_{13}^2 \right)}
= 1 - \frac{3 + \sqrt{5}}{8 \left( 1 - s_{13}^2 \right)}.
\ee
We use the phenomenological data of ref.~\cite{forero}
(see refs.~\cite{fogli,schwetz}
for alternative phenomenological fits to the data).
Using as input the $3 \sigma$ range given therein for $s_{13}^2$,
namely $s_{13}^2 \in \left[ 0.017,\, 0.033 \right]$,
we depict eq.~(\ref{s12sq}) in fig.~\ref{fig:s12sq}.
The prediction for $s_{12}^2$ is in very good agreement
with the experimental value,
$s_{12}^2 \in \left[ 0.303,\, 0.336 \right]$ at $1 \sigma$.

\begin{figure}
\begin{center}
\epsfig{file=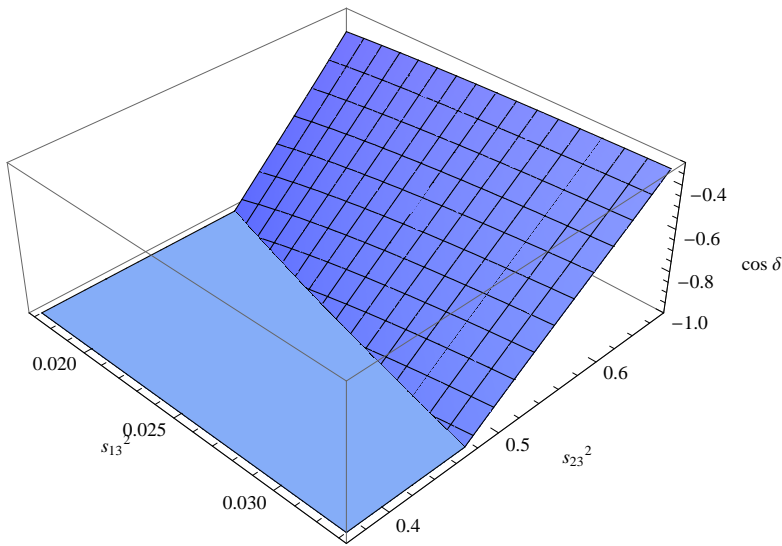,width=0.75\textwidth}
\end{center}
\caption{$\cos{\delta}$ as function of the mixing angles.
\label{fig:cosd}}
\end{figure}

We are able to express $\cos \delta$ in terms of $s_{13}^2$ and $s_{23}^2$
by using eq.~(\ref{pred2}).
The expression is
\be
\cos \delta = \frac{- 2
+ \left( 5 + \sqrt{5} \right) s_{13}^2
+ \left( 5 - \sqrt{5} \right) s_{23}^2
- \left( 11 + \sqrt{5} \right) s_{23}^2 s_{13}^2}
{\left( 1 + \sqrt{5} \right)
\sqrt{2 s_{13}^2 s_{23}^2 \left( 1 - s_{23}^2 \right)}\,
\sqrt{5 - \sqrt{5} - 8 s_{13}^2}
}.
\ee
This is illustrated in fig.~\ref{fig:cosd},
wherein
$\cos \delta$ varies from about $-0.33$ to $-1$.
In fig.~\ref{fig:cosd},
the flat region indicates values of $\cos \delta$ smaller than $-1$;
this means that the model predicts $s_{23}^2$ to be
larger than about 0.48---this is disfavoured by the fit of ref.~\cite{fogli}
but allowed by the other two phenomenological fits.
If future measurements clearly reveal a smaller value of $s_{23}^2$,
then this implementation of our model will be ruled out.

\begin{figure}
\begin{center}
\epsfig{file=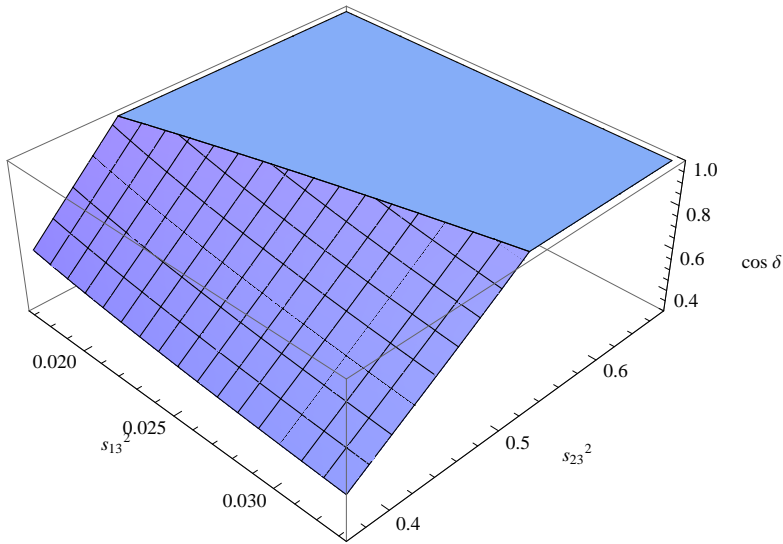,width=0.75\textwidth}
\end{center}
\caption{$\cos{\delta}$ as function of $s_{13}^2$ and $s_{23}^2$
for the alternative realization of the model
where the second and third rows of $U$ are interchanged.
\label{fig:cosd2}}
\end{figure}

We may alternatively choose the option where the eigenvalues of $M_\ell$
that correspond to the masses of the $\mu$ and the $\tau$ charged leptons
are interchanged.
This means swapping $m_\mu$ and $m_\tau$ in eqs.~(\ref{mell});
of course this requires a different set of parameters
to fit the charged-lepton masses.
In this case,
we must swap the second and third rows of the lepton mixing matrix,
so that
\be
\left( c_{23}^2 - s_{23}^2 \right) \left( s_{12}^2 - c_{12}^2 s_{13}^2 \right)
+ 4 c_{12} s_{12} c_{23} s_{23} s_{13} \cos{\delta} = \frac{1 - \mu_+^2}{4}
\ee
instead of eq.~(\ref{pred2}).
In this case the prediction for $\delta$ will be
\be
\cos{\delta} = \frac{- 3 + \sqrt{5} + 6 s_{13}^2
+ \left( 5 - \sqrt{5} \right) s_{23}^2
- \left( 11 + \sqrt{5} \right) s_{23}^2 s_{13}^2}
{\left( 1 + \sqrt{5} \right)
\sqrt{2 s_{13}^2 s_{23}^2 \left( 1 - s_{23}^2 \right)}\,
\sqrt{5 - \sqrt{5} - 8 s_{13}^2}
}.
\ee
This is shown in fig.~\ref{fig:cosd2}.
One sees that in this case $\cos{\delta}$ is positive
and lies between $\sim 0.47$ and $1$,
while $\theta_{23}$ is usually in the first octant,
with $s_{23}^2$ smaller than about $0.51$.

\section{Conclusions}
\label{sec:conclusions}

We have considered in this paper a lepton mixing scheme
arising from an $A_5$ flavour symmetry.
In our model,
the VEVs of the required $A_5$ multiplets
are aligned in a self-consistent fashion
and the neutrino and charged-lepton sectors are separated
through an auxiliary $\mathbbm{Z}_3 \times \mathbbm{Z}_3$ symmetry.

The model predicts two relations
among the parameters of the lepton mixing matrix.
Using the observed $\theta_{13}$ as input,
the solar mixing angle is predicted in good agreement with experiment.
Depending on the assignment of the lepton flavours,
the model either predicts $s_{23}^2 > 0.48$
and negative $\cos{\delta}$ between $-0.33$ and $-1$,
or $s_{23}^2 < 0.51$ and positive $\cos{\delta}$ between $0.47$ and $1$ .

\paragraph{Acknowledgements:} The authors thank Patrick O.\ Ludl
for deriving the Clebsch--Gordan coefficients of $A_5$.
The work of IdMV is supported by the Swiss National Science Foundation.
The work of LL is supported by the Portuguese
\textit{Funda\c c\~ao para a Ci\^encia e a Tecnologia}\/
through the projects PEst-OE-FIS-UI0777-2013,
CERN/FP/123580/2011,
and PTDC/FIS-NUC/0548-2012.

\vspace*{5mm}

\begin{appendix}

\setcounter{equation}{0}
\renewcommand{\theequation}{A\arabic{equation}}

\section{The group $A_5$}
\label{apa}

In suitable bases,
the $\mathbf{3}_1$ and the $\mathbf{5}$ are given as
\be
\label{31}
\begin{array}{c}
{\displaystyle
\mathbf{3}_1: \quad
a \to A_3 \equiv \mathrm{diag} \left( -1,\ -1,\ 1 \right),
}
\\*[2mm]
{\displaystyle
b \to B_3 \equiv \left( \begin{array}{ccc} 0&1&0\\0&0&1\\1&0&0
\end{array} \right),
}
\\*[5mm]
{\displaystyle
c \to C_3 \equiv
\frac{1}{2} \left( \begin{array}{cccc}
\mu_+ & -1 & \mu_- \\
-1 & \mu_- & \mu_+ \\
\mu_- & \mu_+ & -1
\end{array} \right)
};
\end{array}
\ee
%
%
\be
\label{5}
\begin{array}{c}
{\displaystyle
\mathbf{5}:
\quad
a \to A_5 \equiv \mathrm{diag} \left( 1,\ -1,\ -1,\ 1,\ 1 \right),
}
\\*[2mm]
\displaystyle{
b \to B_5 \equiv \frac{1}{2} \left( \begin{array}{ccccc}
0 & 2 & 0 & 0 & 0 \\
0 & 0 & 2 & 0 & 0 \\
2 & 0 & 0 & 0 & 0 \\
0 & 0 & 0 & - 1 & - \sqrt{3} \\
0 & 0 & 0 & \sqrt{3} & - 1
\end{array} \right),
}
\\*[5mm]
\displaystyle{
c \to C_5 \equiv
\frac{1}{4} \left( \begin{array}{ccccc}
0 & 2 & 2 & 2 \sqrt{2} & 0 \\
2 & 2 & 0 & - \sqrt{2} & \sqrt{6} \\
2 & 0 & 2 & - \sqrt{2} & - \sqrt{6} \\
2 \sqrt{2} & - \sqrt{2} & - \sqrt{2} & 2 & 0 \\
0 & \sqrt{6} & - \sqrt{6} & 0 & - 2
\end{array} \right).
}
\end{array}
\ee
%
%
%

Since $c^2 = e$,
$c$ has eigenvalues $+1$ and $-1$.
Because $\mathrm{tr}\ C_3 = -1$,
$C_3$ has a twice degenerate eigenvalue $-1$
and a non-degenerate eigenvalue $+1$;
the normalized eigenvector corresponding to the latter is the vector $v$ in eq.(\ref{vmu}).
%
%
Because $\mathrm{tr}\ C_5 = 1$,
$C_5$ has a twice degenerate eigenvalue $-1$
and a thrice degenerate eigenvalue $+1$;
three orthonormal eigenvectors spanning the subspace
corresponding to the latter eigenvalue are
\be
\label{c5}
\frac{1}{\sqrt{3}} \left( \begin{array}{c}
1 \\ 1 \\ 1 \\ 0 \\ 0 \end{array} \right),
\quad
\frac{1}{4} \left( \begin{array}{c}
0 \\ \sqrt{6} \\ -\sqrt{6} \\ 0 \\ 2 \end{array} \right),
\quad \mathrm{and} \quad
\frac{1}{2 \sqrt{6}} \left( \begin{array}{c}
2 \\ -1 \\ -1 \\ 3 \sqrt{2} \\ 0 \end{array} \right).
\ee

The product $\mathbf{3}_1 \otimes \mathbf{3}_1$ may be decomposed as
\be
\label{31product}
\mathbf{3}_1 \otimes \mathbf{3}_1 =
\left( \mathbf{1} \oplus \mathbf{5} \right)_\mathrm{symmetric}
\oplus \left( \mathbf{3}_1 \right)_\mathrm{antisymmetric}.
\ee
Let $\left( a_1,\, a_2,\, a_3 \right)$ 
and $\left( b_1,\, b_2,\, b_3 \right)$ be two $\mathbf{3}_1$ of $A_5$,
then $\left. \left( \sum_{i=1}^3 a_i b_i \right) \right/ \sqrt{3}$
is $A_5$-invariant and
\be
\label{see31product}
\frac{1}{2 \sqrt{3}} \left( \begin{array}{c}
\sqrt{6} \left( a_1 b_2 + a_2 b_1 \right)
\\
\sqrt{6} \left( a_2 b_3 + a_3 b_2 \right)
\\
\sqrt{6} \left( a_1 b_3 + a_3 b_1 \right)
\\
\sqrt{3} \left( - \mu_+ a_1 b_1 - \mu_- a_2 b_2 - a_3 b_3 \right)
\\
\left( 1 - \mu_- \right) a_1 b_1 + \left( \mu_+ - 1 \right) a_2 b_2
+ \left( \mu_- - \mu_+ \right) a_3 b_3
\end{array} \right)
\ee
is a $\mathbf{5}$ of $A_5$.

The product $\mathbf{5} \otimes \mathbf{5}$ may be decomposed as
\be
\label{5product}
\mathbf{5} \otimes \mathbf{5} =
\left( \mathbf{1} \oplus \mathbf{4} \oplus \mathbf{5} \oplus \mathbf{5}
\right)_\mathrm{symmetric} \oplus \left( \mathbf{3}_1 \oplus \mathbf{3}_2
\oplus \mathbf{4} \right)_\mathrm{antisymmetric}.
\ee
Let $\left( a_1,\, a_2,\, a_3,\, a_4,\, a_5 \right)$ 
and $\left( b_1,\, b_2,\, b_3,\, b_4,\, b_5 \right)$ be two $\mathbf{5}$
of the IG.
Then,
there is an $A_5$-invariant
$\left. \left( \sum_{i=1}^5 a_i b_i \right) \right/ \sqrt{5}$ and
\be
\label{see5product}
\begin{array}{c}
{\displaystyle
\frac{1}{2 \sqrt{2}} \left( \begin{array}{c}
2 \left( a_1 b_5 + a_5 b_1 \right)
\\
- a_2 b_5 - a_5 b_2 + \sqrt{3} \left( a_2 b_4 + a_4 b_2 \right)
\\
- a_3 b_5 - a_5 b_3 - \sqrt{3} \left( a_3 b_4 + a_4 b_3 \right)
\\
- a_4 b_5 - a_5 b_4 + \sqrt{3} \left( a_2 b_2 - a_3 b_3 \right)
\\
2 a_1 b_1 - a_2 b_2 - a_3 b_3 - a_4 b_4 + a_5 b_5
\end{array} \right),
}
\\*[5mm]
{\displaystyle
\frac{1}{2 \sqrt{6}} \left( \begin{array}{c}
2 \left( a_1 b_4 + a_4 b_1 \right)
+ 2 \sqrt{2} \left( a_2 b_3 + a_3 b_2 \right)
\\
- a_2 b_4 - a_4 b_2 - \sqrt{3} \left( a_2 b_5 + a_5 b_2 \right)
+ 2 \sqrt{2} \left( a_1 b_3 + a_3 b_1 \right)
\\
- a_3 b_4 - a_4 b_3 + \sqrt{3} \left( a_3 b_5 + a_5 b_3 \right)
+ 2 \sqrt{2} \left( a_1 b_2 + a_2 b_1 \right)
\\
2 a_1 b_1 - a_2 b_2 - a_3 b_3 + 3 \left( a_4 b_4 - a_5 b_5 \right)
\\
\sqrt{3} \left( a_3 b_3 - a_2 b_2 \right) - 3 \left( a_4 b_5 + a_5 b_4 \right)
\end{array} \right)
}
\end{array}
\ee
are two $\mathbf{5}$ of the IG.

\end{appendix}

\end{document}